\documentstyle[12pt,pic98pro,epsfig,wrapfig]{article}
\begin{document}
\title{ 
COSMIC VERY HIGH-ENERGY $\gamma$-RAYS
}
\author{
R.~Plaga                \\
{\em Max-Planck-Institut f\"ur Physik
(Werner Heisenberg Institut)}, 80805 M\"unchen, Germany}
\maketitle
\baselineskip=14.5pt
\begin{abstract}
The article gives a brief overview, aimed at nonspecialists, 
about the goals and
selected recent results of the detection of
very-high energy $\gamma$-rays (energies
above 100 GeV) with ground based
detectors.
The stress is on the physics questions, especially the
origin of Galactic Cosmic Rays and the emission of TeV
$\gamma$-radiation from active galaxies.
Moreover some particle-physics questions which are addressed 
in this area are discussed.        
\end{abstract}
\baselineskip=17pt
\section{Introduction: Very-high energy
$\gamma$-rays and nonthermal processes in the universe}
\subsection{The nonthermal universe}
\begin{wrapfigure}{r}{10cm}
\epsfig{file=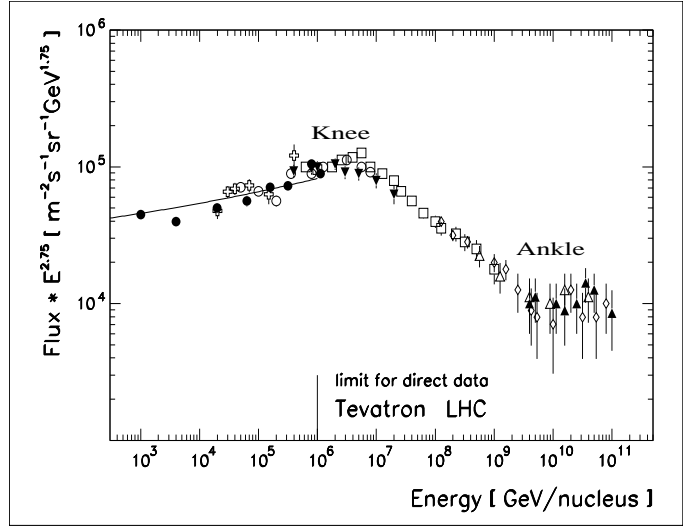,width=9cm,height=7cm,clip=,angle=0}
 \caption{\it Energy spectrum of hadronic cosmic
rays with energies $>$ 300 GeV. The flux measurements
from various experiments (below about 100 TeV 
from direct balloon and satellite
data, above 1000 TeV from ground-based arrays) 
are plotted against the
differential flux, multiplied by energy$^{2.75}$ to remove
the very fast power law decrease in intensity.
Within the errors the spectrum displays a power law behaviour
with three different indices below
and above the ``knee'' (at about 2 $\cdot$ 10$^{15}$ eV) and
above the ``ankle'' at about 3 $\cdot$ 10$^{18}$ eV.
For a detailed explanation of the symbols see
Wiebel\cite{wiebel94}.
      }
    \label{cr_spec} 
\end{wrapfigure}

The general aim of ``very-high energy'' 
(``VHE'', defined as $\gamma$ rays with energies
from about 20 GeV up to about 50 TeV) $\gamma$-ray
astrophysics  is to explore the ``nonthermal
particles in the universe''.  This is a designation for 
those cosmic particles
that have been accelerated to energies at which 
they are no longer in thermal equilibrium
with their surroundings, and for the radiation fields
they produce. 
Discovered at the beginning of this century,
{\it Galactic cosmic rays}
constitute the most striking (but by no means only) evidence
for nonthermal processes. 
They are observed near 
earth as an intense, nearly isotropic, flux 
of mainly protons and other nuclei with energies
ranging from about 100 MeV to at least 
a few times 10$^{20}$ eV. Electrons are a minor component (their intensity, at about
50 GeV energy, is only 1$\%$ of the hadronic flux). 
The hadronic energy spectrum is 
displayed in fig.\ref{cr_spec}.
The total number of cosmic-ray particles, compared to 
thermal interstellar particles, is very small.
Their total energy density, however, 
(about 0.5 eV/cm$^{3}$ near earth) is
comparable to other constituents of our
Galaxy like e.g. interstellar magnetic fields and turbulent
motion in the interstellar medium. 
Nonthermal particles are therefore
likely to play an important role in the general dynamics 
of clusters of galaxies, our Galaxy 
and the objects which accelerate particles.
\\
The basic questions for an understanding of the
nonthermal universe are: {\bf Where, and with what mechanisms
are particles accelerated in the universe?}
While there are experimental clues concerning the first question 
for {\it electrons} (from radio astronomy, 
X-ray and VHE $\gamma$-rays, to be discussed below)
there is as yet scant unambigious experimental evidence as to
the origin of nonthermal {\it hadrons}, which seem to
dominate the total energy density, at least in 
in the special case of Galactic cosmic rays. 
As an example for our
lack of knowledge: for the energies 
shown in fig.\ref{cr_spec} ($>$ 100 GeV/nucleus)
there is no direct {\it experimental}  evidence {\it whatsoever}
about the local density of hadronic cosmic rays
anywhere in the universe except near earth. 
\\
A mechanism that is very likely to play an important
role in the acceleration, is the ``first order Fermi mechanism''
\cite{acctheo,petheo}.
When macroscopic amounts of matter are ejected 
into interstellar space with 
velocities larger than the sound speed in the ambient 
medium, a shock wave forms. Reflected by inhomogeneities
in the ambient magnetic fields, 
charged particles then have 
a certain probability to repeatedly cross the shock front
and gain energy each time they do so. Extensive
analytical and Monte Carlo studies have shown that
this is a plausible mechanism to convert a nonnegligible
part of the kinetic energy of the shock front into 
the energy of nonthermal
particles. This mechanism predicts 
energy spectra obeying power laws with indices
near $\alpha$=-2 which is in agreement with
observed spectra when additional mechanisms
affecting the spectral form (e.g. energy-loss and propagation
effects) are taken into account.
\\
Theory is not yet able to a priori 
predict the hadron to electron
intensity ratio, however\cite{petheo}.  
The ambiguity in this crucial parameter, 
perhaps the most important single roadblock to
a better understanding of the nonthermal universe, will
reappear the discussion of VHE 
$\gamma$-ray sources below.

\subsection{Ground based VHE $\gamma$ astronomy}
\begin{figure}[htb]
\centering
\epsfig{file=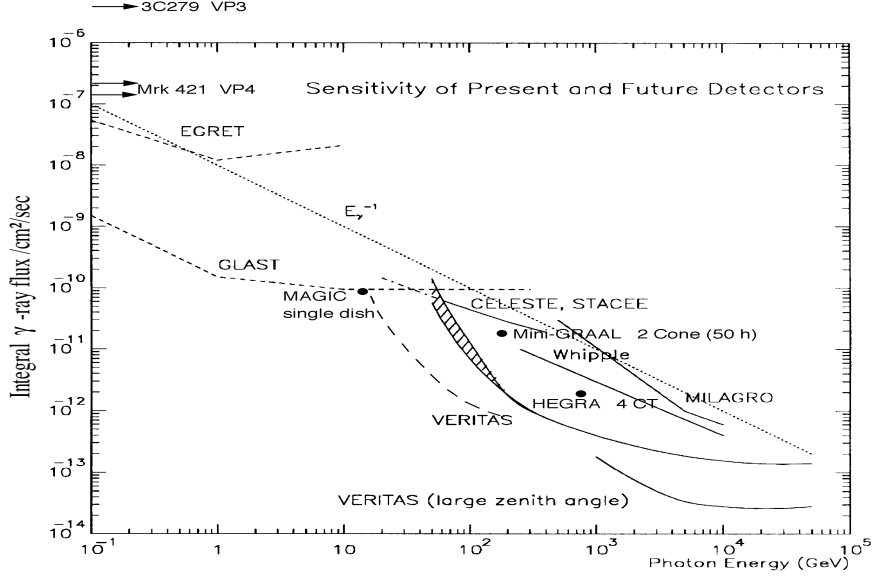,
width=12cm,height=8cm,clip=,angle=0}
  \caption{\it Sensitivities 
of various existing and planned detectors (adapted from VERITAS
proposal\cite{whipplep}),
for high and very-high energy gamma rays for various energies
(full lines) or at the energy threshold (big dot). 
A total measuring time of 50 hours
was assumed  for devices with a small
field of view and 1 year for devices with a field of view
of 1 steradian or larger (these are:
satellites: EGRET (existing) and GLAST
(planned) and the array MILAGRO\cite{milagro}
(under construction)).
WHIPPLE\cite{whipple} and HEGRA\cite{hegra}
are existing telescopes, VERITAS\cite{whipplep}
and MAGIC\cite{magic} proposed large future devices;
HESS\cite{hess}, another multi-telescope 
proposal, is expected to reach
similar sensitivites as VERITAS.
CELESTE, STACEE,
and GRAAL are detectors which are currently being set
up using the large mirror areas at solar power plants\cite{solar}. The dotted
line gives the extrapolation of a typical VHE 
source spectrum from
GeV to VHE energies (active galaxy Mkn 421). The arrows indicate
the intensity of this source during various
viewing periods (``VP'') of the EGRET satellite.}
\label{sens}
\end{figure}

The currently dominating technique to measure cosmic
$\gamma$-ray photons with energies 
from about 200 GeV to about 20 TeV uses
\v{C}erenkov telescopes. The primary gamma ray develops
an electromagnetic cascade in the earth's atmosphere
which emits \v{C}erenkov
light. This light is detected in an optical telescope with
typical mirror areas from about 5 to 70 m$^2$.The \v{C}erenkov light is emitted over an area of
about 20000 m$^2$ and each telescope 
situated whithin this ``light pool'' can detect the shower.
This large detection area allows to reach good counting statistics: 
E.g. the Whipple collaboration detected 
in one observing season (spring '96) about 5000 photons
from the active galaxy Mkn 421 above 300 GeV\cite{whipplep}.
Until the end of the 1980s achievable
sensitivities were seriously degraded
by the fact that the hadronic showers from Galactic cosmic-rays produce
a background which, even for the strongest sources,
is about a factor 100 higher than the signal from photons.
The imaging technique, 
pioneered by the Whipple collaboration\cite{weekes}
allows to reduce this background.
A picture of the shower is registered
in an array of fast photomulipliers in the image plane
of the telescope that detect the
few nanosecond long \v{C}erenkov pulses. 
This technique is still in a state of evolution, the most
advanced camera, presently in
operation by the French CAT collaboration\cite{cat}
has 546 pixels. Another crucial technological advance for
\v{C}erenkov telescopes, 
mostly advanced by the HEGRA collaboration,
is the exploitation of
stereo imaging\cite{hofm}. The same shower is viewed by several
telescopes in coincidence and its full three
dimensional structure can be reconstructed.
Besides advantages like a better
energy and angular resolution, the measurement 
of shower parameters is overdetermined
and thus allows important consistency checks of the
technique.
The capabilites of existing detectors and planned
future arrays of \v{C}erenkov devices is
illustrated in fig.\ref{sens}.  
\subsection{The astrophysical 
production of VHE $\gamma$-rays}
\label{gprod}
Nonthermal particles produce $\gamma$-rays
in the interaction with ambient matter and radiation fields.
As opposed to the particles
themselves, which are charged 
and therefore isotropized in Galactic magnetic fields, $\gamma$-rays point
back to the site of their production. At an acceleration site   
the density of nonthermal particles 
- and therefore the $\gamma$-ray emissivity -  
is expected to be high.
The detection of VHE $\gamma$-ray sources thus
yields clues about cosmic sites  
where particles are accelerated to energies of about
100 GeV and above. 
The two most important $\gamma$-ray production mechanisms
for particles interacting with ambient {\it matter} 
(normally mostly hydrogen (H) are):
\\
{\sl 1. for electrons: relativistic Bremsstrahlung 
(e + H $\rightarrow$ e + H + $\gamma$)}
\\
The energy loss $\delta$E of this mechanism
for electrons with an energy E$_e$ in
``astronomical units'' 
($\rho$ is the ambient interstellar number density):
\\
\begin {equation}
{\delta E \over  Lightyear} \approx 25 keV \cdot {E_e \over TeV}
\cdot {\left({\rho \over {(1/cm^3)}}\right)}
\label{brem}
\end{equation}
One obtains for the typical energy 
$\gamma$-ray energy E$_{\gamma}$
produced by an electron of energy E$_e$: 
E$_{\gamma}$ $\approx$ {$E_e\over 3$}.
\\
{\sl 2. for hadrons: pion production in nuclear interactions
(p + H $\rightarrow$
p + H + $\pi_0$; $\pi_0$ $\rightarrow$
$\gamma \gamma$)}
\\
$\pi_0$ decay produces the 
$\gamma$'s  in this mechanism.
For the energy loss one gets for proton energies
$>$ 1 TeV:
 \begin {equation}
{\delta E \over  Lightyear} \approx 20 keV \cdot
{\left({\rho \over {(1/cm^3)}}\right)}
\label{nuc}
\end{equation}
For the typical energy one gets  
E$_{\gamma}$ $\approx$ {$E_p\over 5$}.
\\
Taking into account the lower mean 
energy and the fact that only a part of the produced
pions decays into $\gamma$'s, it can be seen that 
the nucleonic mechanism is slightly less efficient.
However, if the total density of 
nonthermal hadrons dominates
by a large factor  (as is suggested by their ratio
of about 100 in the Galactic cosmic rays near earth),
this $\gamma$-ray production mechanism can dominate.
\\
For electrons two other mechanisms operating
on ambient {\it radiation fields} are often more
important than Bremsstrahlung:
{\sl 3. Synchrotron radiation on an ambient
cosmic magnetic field B :  
e + B-field $\rightarrow$ e + B-field + $\gamma$}
\\
The energy loss is given as:
\begin {equation}
{\delta E \over  Lightyear} \approx 700 keV \cdot 
{\left({E_e \over TeV}\right)^2} \cdot {\left({B \over 3 \mu G}\right)^2}
\label{syn}
\end{equation}
In the VHE energy range this is more efficent than
the mechanism on matter for densities smaller than about 30 cm$^3$
(i.e. outside dense clouds). The  typical energy
is quite small however:
\begin{equation}
 E_{\gamma} \approx 0.05 
{\left(E_e\over TeV\right)^2 \cdot 
\left({B \over 3 \mu G}\right)^2 eV}
\label{csy}
\end{equation}
\\\begin{figure}[tbh]
\centering
\epsfig{file=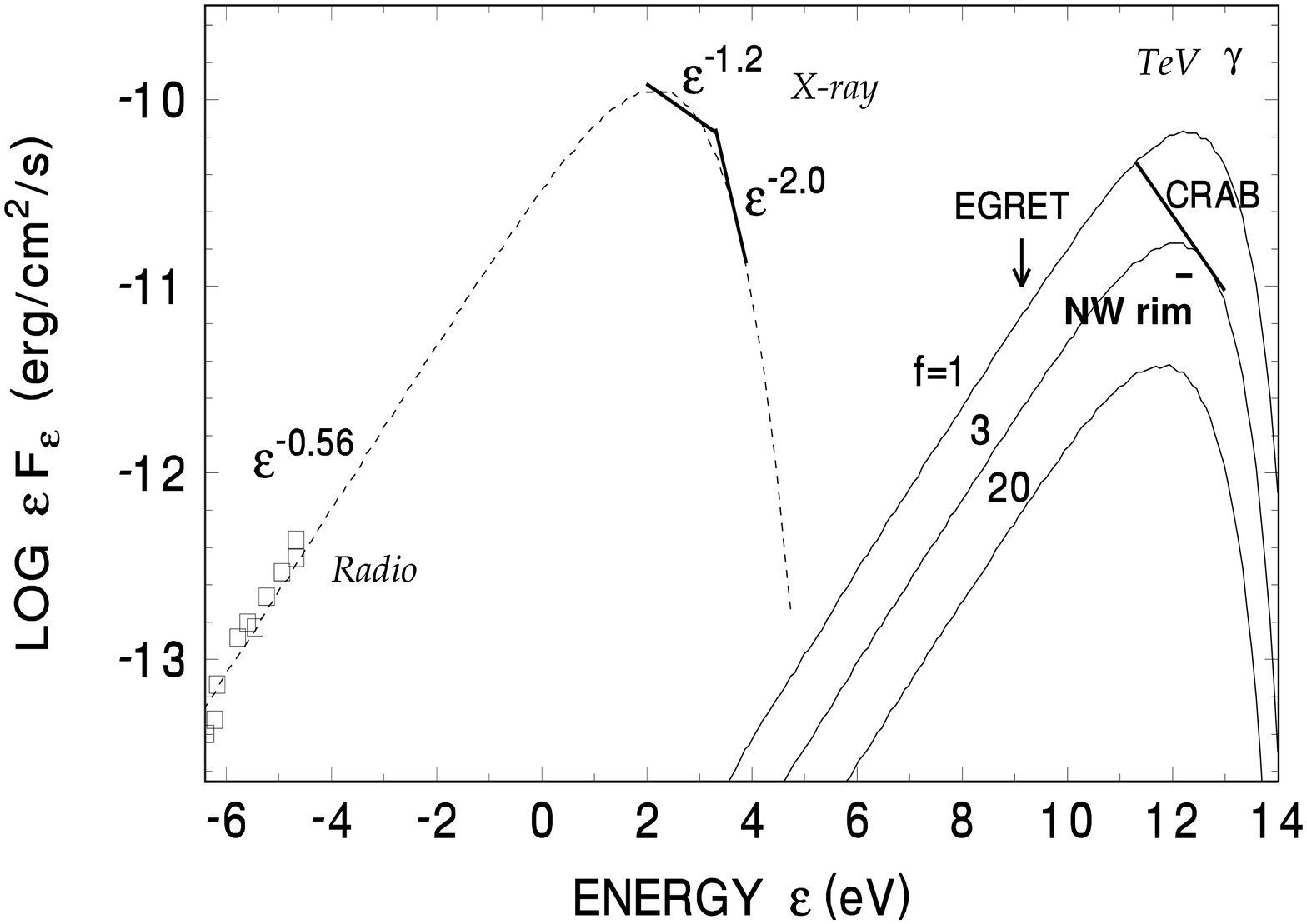,width=12cm,height=8cm,clip=,angle=0}
 \caption{\it The energy spectrum of the supernova remnant 1006
from radio to the TeV $\gamma$-range. The flux units are chosen such
that equal size on the y-axis corresponds to equal total energy
output. This figure was prepared  by Mastichiades and de Jager
\cite{masti} {\bf before} the CANGAROO 
collaboration discovered TeV $\gamma$-rays above 3 and 1.7 TeV from
the NW rim of the remnant. 
This experimental flux is symbolized by the small horizontal bar
which was inserted in the diagram under the assumption of
a spectral energy index of -2 (used by CANGAROO in their
data anlysis). The full lines labeled f=1 etc. are predictions
based on fit to the experimental data at low energies
(dashed lines) for the total remnant, the predictions only for
the NW rim would be about a factor 2-3 lower.  
For comparison the full line labeled
``Crab'' shows the TeV energy spectrum of Crab nebula,
a ``standard candle'' which has been measured by various
ground-based groups over a large energy range.     
      }
    \label{1006fig1} 
\end{figure}
We will see below that photons produced
by this mechanism (typically in the radio to X-ray range)
have probably been observed together with photons produced by
the the following mechanism:
\\
{\sl 4. Inverse Compton scattering on ambient low
energy photons with an energy E$_a$ and an
energy density U
(e.g. the thermal photons from
the cosmological 3 K background radiation
with a typical energy of $2 \cdot 10^{-4} eV$)}:
\\
e + $\gamma$ $\rightarrow$ e +  $\gamma$
\\
with 
\begin {equation}
{\delta E \over  Lightyear} \approx 700 keV \cdot 
{\left({E_e \over TeV}\right)^2} \cdot 
{\left({U \over 0.22 ev/cm^{3}}\right)^2}
\label{ic}
\end{equation}
0.22 eV/cm$^3$ is the energy density equivalent
to a magnetic field of 3 $\mu$G and close to 
the density of the 3 K radiation (0.26 eV/cm$^3$).
For this mechanism the typical energy
rises fast with energy:
\begin {equation}
E_{\gamma} \approx 1.3 
{\left(E_e\over TeV\right)^2 \cdot 
\left({E_a \over {2 \cdot 10^{-4}}}\right)^2 GeV}
\label{cic}
\end{equation}
Consequently at high $\gamma$-ray energies this  mechanism
is often the most important one.
\\
Processes 3. and 4. operate also with protons, but
only at energies larger by a factor (m$_p/m_e)^2$
$\approx$ 3 $\cdot$ 10$^6$, where for 
typical nonthermal spectra
integral flux  has typically fallen at least by a similar factor. 
\\
\section{Supernova remnants, pulsars and the origin of Cosmic Rays}
Cosmic rays with energies below the ``knee'' (see fig.\ref{cr_spec})
are widely
believed to be mainly accelerated in ``supernova remnants'' 
(``SNR's'', the debris, consisting of hot thermal matter
and a shock wave running into the interstellar medium, 
of supernova (``SN'') explosions) in our Galaxy.
The following ``standard scenario''
for the origin of 
Galactic cosmic rays is plausible and backed
by extensive theoretical work on acceleration
of particles in the shock waves induced by
the supernova explosion\cite{acctheo}. It is known that
about every 30 years a supernova explodes in our
Galaxy. From the experimental determination of 
spallation products
in the cosmic rays it is deduced that the mean lifetime of cosmic
rays, confined by magnetic fields to the
Galaxy, is about 3 $\cdot$ 10$^{7}$ years.
From this one can determine that in order to sustain
the above mentioned observed energy density of cosmic
rays, about 10 $\%$ of the kinetic energy released
in a typical SN (10$^{51}$ ergs) has to be converted into 
nonthermal particles. If this scenario is to explain
the hadron to electron ratio observed near earth in a simple
manner, the hadron to electron ratio in the accelerated particles
has to be about 100.   
\\
SN explosions seem to be the {\it only} 
Galactic phenomena
which release enough kinetic energy to sustain the
observed cosmic-ray density against losses
\footnote{The question if ``$\gamma$-ray bursts''\cite{bignami} could
be important in this respect can not be reliably
answered at present.}.    
In particular pulsars, rapidly rotating magnetized
neutron stars, fall short of the required kinetic energy
by about an order of magnitude\cite{acctheo}.
Direct evidence for the acceleration
of cosmic rays in SNR's is necessary to confirm
this scenario.
\\
In  1995 the ASCA satellite observed X-ray radiation
from the SNR 1006, the product of a SN explosion in the year 
A.D. 1006 in a distance of about 1.8 kpc\cite{koyama}. 
Based on the measured spectrum
and emission region at the rim of the SNR, 
this radiation was interpreted as being due
to synchrotron radiation from electrons 
recently accelerated in the
remnant to  energies of up to 100 TeV, which leads to
energies for synchrotron radiation in the X-ray range
(eq.(\ref{csy})). 
If this interpretation is correct, a glance at
eq.(\ref{syn}) and (\ref{ic}) 
makes clear that 
the SNR should also be a copious emitter of 
inverse Compton $\gamma$-rays.
\begin{figure}[tbh]
\centering
\epsfig{file=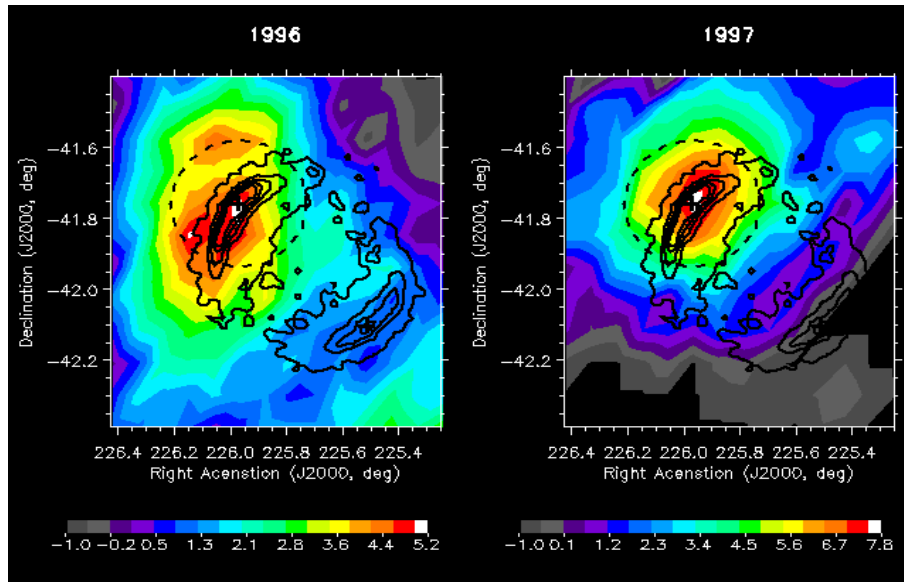,width=12cm,height=8cm,clip=,angle=0}
 \caption{\it 
      Sky map of the remnant of  the supernova explosion of A.D.1006
      (from Tanimori et al.\cite{cang}), a shell with a diameter of
      about 20 parsecs in a distance of about 1.8 kpc.
      The grey scale indicates the
      intensity of $\gamma$ radiation above 3 TeV (for 1996) and 1.7 
      TeV (for 1997) in units of the signal to shot background noise ratio.
       The full lines indicate the intensity of  X-ray
      emission (2-10 keV) as measured with the ASCA satellite\cite{koyama}.
      The remnant's shell  emits nonthermal x-rays mainly from
      two rims in the south west and north east-
      the TeV radiation can be seen to originate only in
      the latter. The dotted circle is the point spread resolution function
       of the CANGAROO \v{C}erenkov telescope. The crosses are
      the directions of maximum x-ray intensity.}
    \label{1006fig2} 
\end{figure}
\begin{figure}[hbt]
\centering
\epsfig{file=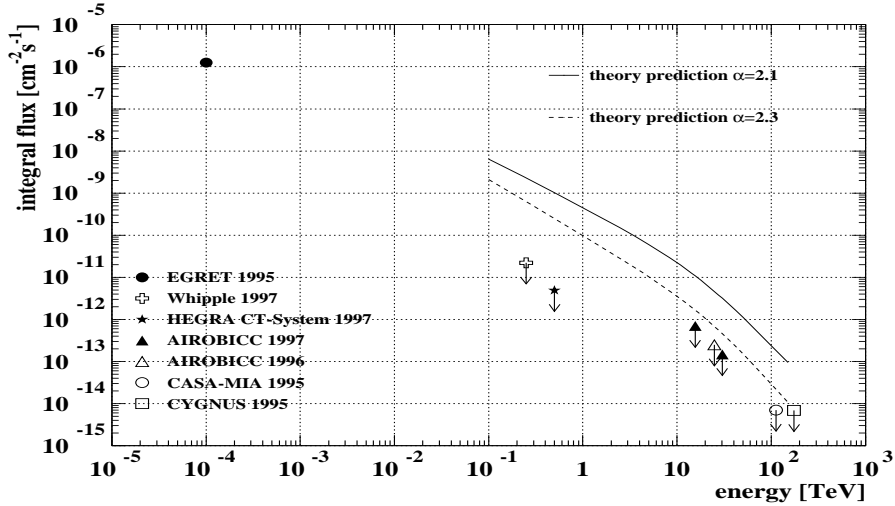,width=12cm,height=7cm,clip=,angle=0}
 \caption{\it The predicted flux in the ``standard scenario''
from the SNR G78.2+2.1,
assuming that the associated molecular cloud
Cong 8 interacts with the nonthermal particles in the SNR and 
no additional complications (see text) is indicated as
the full and dotted line for two different spectral indices
spanning the theoretically plausible range. The upper 
observational  limits from various ground based $\gamma$-ray
detectors are also indicated. The theoretical calculation 
took an expected cutoff in the energies of accelerated
particles due to the finite age of the remnant (ca. 5000 years)
above 100 TeV into account.    
      }
    \label{chrispic} 
\end{figure}
Their typical energy can be inferred from eq.(\ref{cic})
to lie in the TeV region for electron energies above 10 TeV. 
Mastichiadis and de Jager\cite{masti} fitted the X-ray
and further radio data (which is also believed to have a 
synchrotron origin) in a model with two free
parameters: the magnetic fields strength in the remnant
and the spectral index of the power law characterizing
the energy spectrum of the remnant (see fig.\ref{1006fig1}).
This fit led to quite plausible model parameters 
($\alpha$=$-$2.1 as expected in first order Fermi acceleration,
and a maximal energy of about 20 TeV which is in the theoretically
expected range for young SNRs) and {\it predicted}
the TeV $\gamma$-ray flux as a function
of one parameter ``f'',  which is related to 
acceleration  efficiency and the magnetic field
strength in the remnant.        
\\
The Australian-Japanese  
CANGAROO collaboration, which operates a
\v{C}erenkov telescope  with 3.8 m diameter in 
Woomera, Australia, observed this remnant      
in 1996 and 1997 and found evidence for $\gamma$ radiation     
above 1.7 TeV from the direction of the so called ``NW rim''
of this remnant\cite{cang}(fig.\ref{1006fig2}). The measured flux
is consistent with the one predicted by Mastichiades 
and de Jager (see fig.\ref{1006fig1}). With this measurement 
of the inverse Compton component all parameters
of their model are fixed. The magnetic field strength in
the remnant is about 6 $\mu$G and about 1 $\%$ 
of the kinetic energy  in this SN explosion was 
apparently expended to accelerate electrons.
The fact that the ``SW rim'', which is also bright in X-rays,
was not seen by CANGAROO is somewhat surprising and needs
further studies. If confirmed by further measurements,
this detection of a shell type SNR (with no central
neutron star)
strongly supports the general idea that 
the Galactic cosmic-ray electrons are mainly produced 
in SNR's.        
\\
Because of the theoretical difficulties in predicting
the hadron/electron ratio in Fermi acceleration,
mentioned in the introduction,
independent experimental evidence for 
the acceleration of hadrons in the required amount in
SNR's is necessary.  The predicted VHE $\gamma$-ray 
fluxes for ``average'' SNR's
turn out to be at the very limit of the capability of
present detectors\cite{flupred} and 
interest concentrated first on remnants
which have an  ``atypically'' high density of ambient 
matter so that the signal due to $\pi_0$ production is
large (see section \ref{gprod}). A prime example
is the SNR G78.2+2.1 in a distance of about 
1.5 kpc\cite{aha94}. There is some evidence 
(from radio and infrared observations) that this remnant
is interacting with a dense molecular cloud (``Cong 8'')
which would act as a ``target'' for $\pi_0$ production.
It is not completely certain that the cloud is not just
superposed on the sky
in front of the SNR, moreover it could e.g. be that
a strong magnetic field  keeps charged cosmic-rays out of it
or that the high matter density inhibits the acceleration
process\cite{acctheo}. Leaving these 
potential complications aside, the 
expected VHE $\gamma$-ray luminosity of such
remnants in the standard ``scenario'' (i.e. that about 10$\%$
of the kinetic explosion energy is expended in acceleration 
of hadrons) is very high (see fig.\ref{chrispic}). 
\\
This SNR was therefore one of the most
extensively  observed ``potential''
VHE $\gamma$-ray source, namely by the
Whipple telescope in (Arizona,
(energy threshold 300 GeV), the HEGRA array of telescopes
(Canary Islands,
energy threshold 500 GeV) and the HEGRA / AIROBICC
(Canary Islands, 12 TeV), Cygnus (New Mexico, 200 TeV)
and CASA arrays (Utah, 100 TeV).   
No evidence for any $\gamma$-ray emission above 300 GeV
was obtained by any experiment, some upper limits being
more than an order of magnitude below the ``naive''
prediction. Even if no interaction with the cloud is assumed
the flux expected under simple assumptions is about
a factor 7 higher than some of the upper limits\cite{acctheo},
because the ambient interstellar matter density is relatively
high at the  Galactic location of the SNR (the ``Cygnus'' region).
Somewhat less restrictive results have been obtained
from the similar case, SNR ``IC 443''.
\\
Unfortunately these negative results are not as
revealing as a detection of VHE $\gamma$-radiation
would have been. As mentioned above, possibilities remain 
to explain the results for these ``atypical'' remnants 
without bringing the ``standard scenario'' into difficulties. 
The ``standard scenario'' remains the most plausible explanation 
 for the origin of the bulk of cosmic rays with energies
below the ``knee'', but there is as yet 
no unequivocal experimental evidence
for it and thus still room for even qualitatively different
explanations (like e.g. an extragalactical origin
of the hadronic component\cite{plaga97}). 
The search for $\gamma$-rays from
$\pi_0$ decay from ``typical'' SNRs like e.g.
``Tycho A'' is the ``experimentum crucis'' for
the ``standard scenario''.
\\
\begin{figure}[tbh]
\centering
\epsfig{file=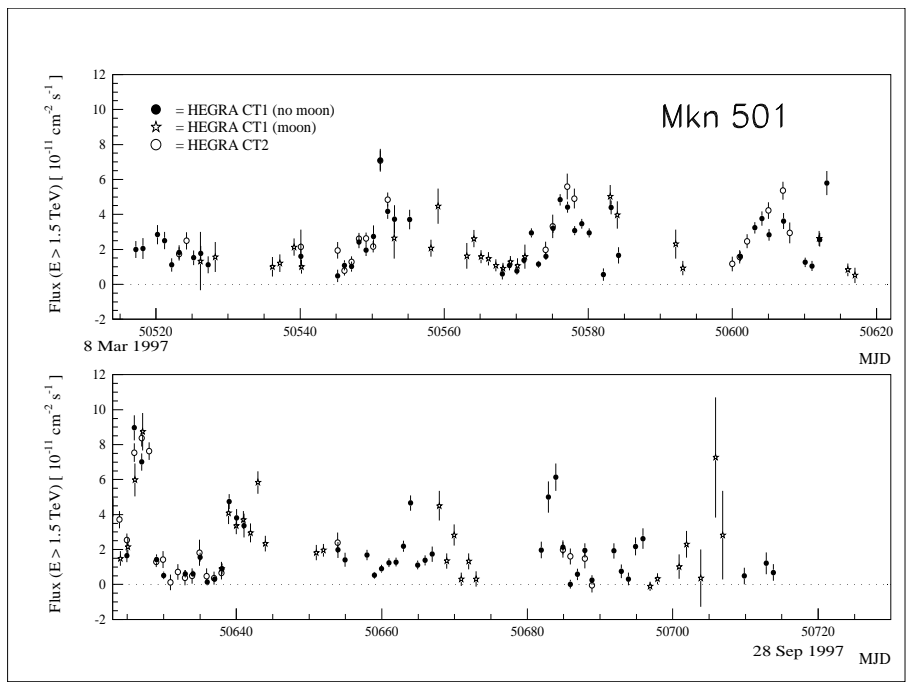,width=12cm,height=9cm,clip=,angle=0}
 \caption{\it Daily integral
flux values of VHE $\gamma$-radiation (energy 
threshold 1.5 TeV) measured with two independent HEGRA
\v{C}erenkov telescopes during 1997. The flux of the the
Crab nebula (a ``reference source'') above this energy  
is 0.8 $\cdot$ 10$^{-11}$ cm$^{-2}$ sec$^{-1}$. 
The data points in moon-lit nights (stars) were taken with 
a reduced operating gain and renormalized to the other 
measurements.}
\label{lightc501} 
\end{figure}
There are three confirmed VHE $\gamma$-ray sources
which  are powered by pulsars
(Crab nebula, Pulsar B1706-44 and Vela pulsar)
\cite{kifune}. As mentioned
above, these are probably not directly relevant for
the origin of the main part of cosmic rays. 
These detections allow to learn about processes
in the magnetosphere of neutron stars\cite{harding}.
Moreover the Crab nebula, a steady, bright VHE source,
has now assumed the role of a ``standard candle'' of TeV photons
(fluxes being somtimes quoted in ``Crab units'').
\section{Extragalactic sources: Active galactic nuclei} 
\begin{wrapfigure}{r}{9cm}
\centering
\epsfig{file=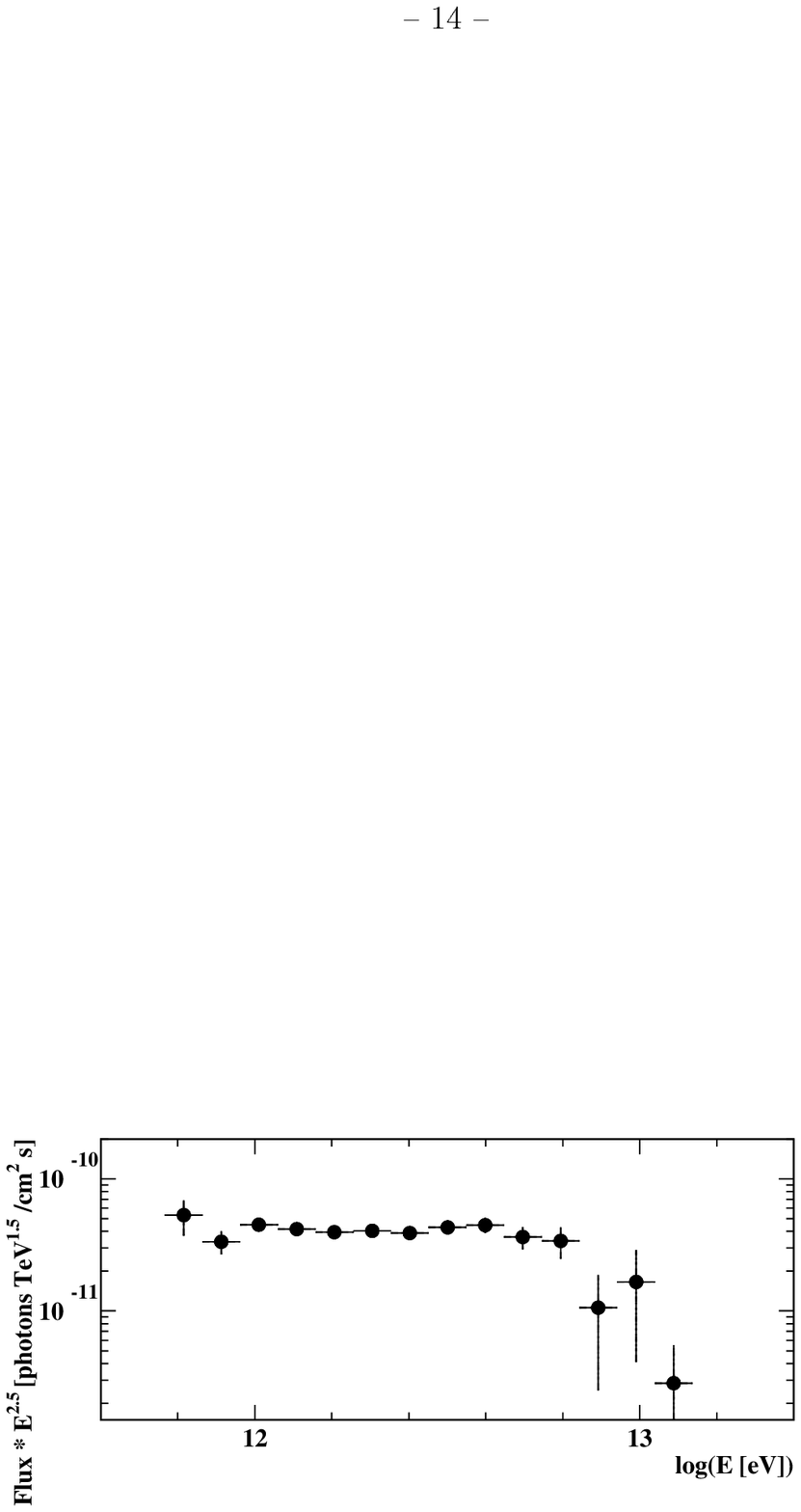,width=8cm,height=7cm,clip=,angle=0}
 \caption{\it Spectrum of the blazar Mkn 501 above
0.6 TeV measured in coincidence in three telescopes
by the Japanese ``Telescope Array''
collaboration\cite{ta}. The spectrum below
about 5 TeV s well aproximated by a power law
with an index of -2.5, quite similar to the 
one of the spectrum of charged-cosmic rays 
near earth of -2.7 (see fig.(\ref{cr_spec}).}
\label{tesh} 
\end{wrapfigure}

Two extragalactic VHE $\gamma$-ray sources,
discovered by the Whipple collaboration,  have
been extensively studied since 1991: the so called
``blazars'' Mkn 421 and 501\cite{mkn}. The most striking observational 
characteristic of these sources is their extreme flux
variability. Mkn 501 was observed in 1995 and 1996
by the Whipple and HEGRA collaborations and
had an intensity of about 8$\%$ of the Crab in 1995.
The intensity rose somewhat in 1996 and entered
a variable ``high state'' in 1997 with maximal fluxes 
up to 10 times the Crab flux, 
more than 100 times the discovery flux,
making it by far the
brightest known VHE source in the sky (see fig.\ref{lightc501}).
In 1998 the source has returned to flux levels of about 40 $\%$ the Crab flux\cite{kestel}. 
\\
The VHE radiation from Mkn 421 is variable on even
smaller time scales down to 15 minutes\cite{wnat}.
The spectrum of Mkn 501 during 1997 has been measured
by the CAT, HEGRA, Whipple and Telescope Array
collaboration between 0.2 and 30 TeV;
as an example the Telescope Array\cite{ta} spectrum
is shown in fig.\ref{tesh}.
Like the the spectrum of Mkn 421 it seems to be reasonably
approximated by a power law below 5 TeV and 
has an exponential cutoff at higher energies, similar results
have been  recently announced by the Whipple and
HEGRA collaboration.
\\
Mkn 421 and 501 are elliptical
galaxies both in a distance of about 150 Mpc which 
probably harbor black holes with masses on the order
of 10$^{6-7}$ solar masses in their centers.
This black hole accretes ambient matter which 
emits radiation before it vanishes in the black hole, thus
forming an ``active galactic nucleus''(AGN). 
The black hole 
powers a jet, which streams out with Lorentz
factors on the order of 10 at a right angle to 
the accretion disk. ``Blazars'' are thought to correspond
to AGNs  which jet is directed into our line of sight. 
This ejection of  matter 
\footnote{analogous
to the case of the SN ejected matter in the previous
section} 
is thought to lead to shock fronts and ensuing
first order particle acceleration at distances on the order
of 10-100 astronomical units from the central black hole.  
VHE $\gamma$-rays could be produced
both by electrons and hadrons. Because of the small
ambient matter density, in {\it both} cases photons
(either ambient or produced via synchrotron radiation
of the nonthermal particles) are expected to constitute
the ``target''.
\\
Fig.{\ref{pian}  illustrates models from Pian et al.\cite{pianp}
in which the total
electromagnetic spectrum of blazars is explained in a way
which bears a qualtitative semblance to the  
model of Mastichiadis, de Jager 
for SN 1006 (fig.\ref{1006fig1}): in both
cases the observed X-rays / TeV $\gamma$-rays
are explained as being due to synchrotron /   inverse
Compton radiation from the same population of
freshly accelerated nonthermal electrons.
It was observed  in 1997 that the intensity variations 
displayed in  VHE $\gamma$ radiation, X-rays and 
optical radiation of Mkn 501 were simultaneous within the measurement 
errors\cite{whipvar}. This supports the idea 
of one nonthermal population which
explains the whole electromagnetic spectrum.
Models based on an hadronic origin of the TeV
$\gamma$-rays can also explain all the observed
properties, however\cite{mannh}.  
Because electrons loose energy more readily
than hadrons (for the same reasons than their
higher efficiency in $\gamma$-ray production), it
is a relatively firm prediction that they cannot
be accelerated to energies much above 10 TeV in the
these blazars\footnote{The fact that this acceleration limit is
similar in energy to the one in SN 1006
is purely coincidental.}. Hadrons, 
 on the other hand, {\it have} to be accelerated
to much higher energies to be viable candidates
as we have seen. So the behaviour   
of the {\it source} spectra around 10 TeV will be
a crucial  experimental fact for the decision of
the basic production mechanism\footnote{
Other 
interesting arguments in favor of either one mechanism
have been advanced\cite{buc}, but are not conclusive in my
opinion}.
\begin{figure}[hbt]
\centering
\epsfig{file=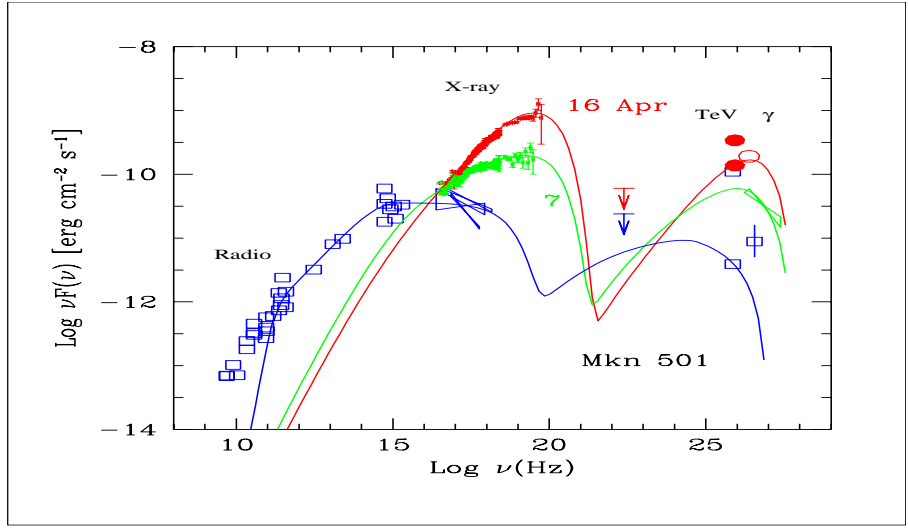,width=12cm,height=7cm,clip=,angle=0}
 \caption{\it Electromagnetic spectrum of the blazar 
Mkn 501 as measured by various experiments\cite{pianp} (see 
Pian et al. for detailed references;
open symbols at left: archival radio, millimeter and X-ray data;  
small black symbols: X-ray flux, Beppo-SAX
satellite on the two
indicated days; upper limits at 0.1 GeV 
(2.4 $\cdot$ 10$^{22}$ Hz):
EGRET collaboration non-coincident;
filled circles:
Whipple collaboration fluxes on these days;
open circle: HEGRA collaboration on April 13;
open squares: 1995/96 data from Whipple; spectral fit:
15-20 March 1997 HEGRA collaboration. The full lines
are theoretical fits for the quiescent, 7th and 16th of April
flux levels (see text).}
\label{pian} 
\end{figure}
The cutoff which seems to be indicated above 5 TeV
(see fig.\ref{tesh})
could be due to such an acceleration limit but also,
{\it alternatively} 
due to intergalactic absorption
(see next section)\cite{steckerir}
with a source spectrum extending to much higher 
energies,
a phenomenon completely unrelated
to the source properties.  
In this connection a prime experimental task is to find
more blazars at different distances from our Galaxy.
Up to now in
searches for further blazars, conducted
by the Whipple and HEGRA  collaborations, 
only the blazar ``1 ES 2344+514'' was uncovered by former group
above 300 GeV during one outburst\cite{1ES}.
%
\section{Fundamental physics with cosmic accelerators}
Experiments concerning
interactions of hadronic cosmic rays in the 
earth atmosphere stood at the
cradle of high-energy physics in the 1940's and 50's.
A similar area of fundamental research is gaining importance,
replacing charged cosmic rays with very-high
energy $\gamma$-rays and the earth atmosphere with
radiation fields in intergalactic space.    
I discuss a recent particle-physics result
using this technique and
two other proposals for more advanced studies. 
\\
The most important cosmological fields of thermal
photons ``between the galaxies'' are
the 3 K background  from the big bang 
and star light with higher Planck temperatures,
mainly from an era of high star formation 
in the early universe during a time 
corresponding to redshift z between of about 
1 and 3\cite{steckerir}
or from even earlier times\cite{primack}.
\begin{wrapfigure}{l}{8cm}
\centering
\epsfig{file=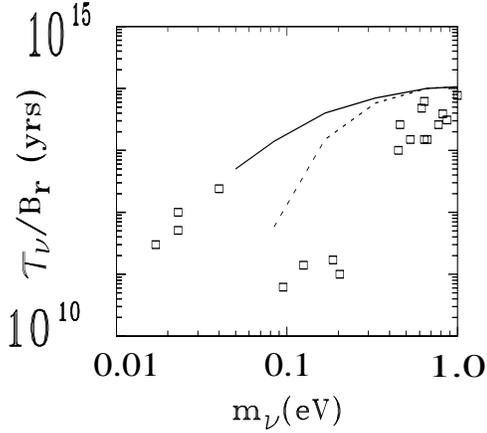,width=7cm,height=6cm,clip=,angle=0}
 \caption{\it Limits on the ratio of radiative lifetime 
to branching ratio of any massive neutrino
species decaying into a 
much lighter neutrino state and a photon
(from Biller et al.\cite{biller}). The full 
(nominal result) and dotted (result allowing for systematic
errors in experimental energy scale)
lines are the lower lifetime limits derived from VHE
observations. The dots are previous lower limits 
derived with other methods
to constrain the infrared background. The standard-model
radiative lifetime of a massive Dirac neutrino radiatively decaying
via flavor mixing is 
10$^{36.5}$ m$_{\nu}^{-5}$ sin$^{-2}$($\theta$) years  
($\theta$ is the generational mixing angle), but there are models
beyond the standard 
model that are ruled out with the given neutrino masses by these 
limits\cite{roos}.
}
    \label{billerfig3} 
\end{wrapfigure}
VHE $\gamma$ rays in the TeV range
are absorbed in 
the reaction $\gamma$(TeV)+$\gamma$(IR) $\rightarrow$ e$^{+}$ e$^{-}$.
\\
The fact that TeV $\gamma$-rays
are observed 
from AGN's in a distance of about 150 Mpc 
that were not  absorbed by the infrared background
allows to set upper limits on the density of this background
between about 1 and 30 $\mu$m wavelength (0.03 eV-
1 eV) \cite{biller,magn,fran}
that are lower than experimental limits with any other method.
In turn, this allowed Biller et al.\cite{biller} to set stringent limits on a 
hypothetical radiative decay of primordial neutrinos in a mass
range from 0.05 - 1 eV (fig.\ref{billerfig3}).
The primordial neutrino density is predicted with great confidence 
from the standard big bang theory . A too short
lifetime would lead to an overproduction of
cosmological infrared photons.
\\
It seems likely that
the absorption of TeV photons sets in slightly above 10 TeV 
for the a distance corresponding to the 
known strong TeV sources Mkn 421 and 501\cite{steckerir}.
The pairs produced in the absorption process  will Compton
upscatter  mainly 3 K background photons to somewhat
smaller energies(eq.(\ref{cic})). An ``intergalactic cascade''
(the pendant to an airshower) develops and at low enough
energies the cascade photons will no longer be absorbed.
If the pairs are not deflected by 
intergalactic magnetic fields they reach the oberver from
the same direction and practically the same time as
the ``original'' source photons. If intergalactic fields
do exist, the pairs are deflected.
This has led to two complementary proposals for fundamental
studies with VHE photons:
\\
1. Aharonian et al.\cite{ahapair}
considered the case of a relatively strong 
intergalactic magnetic field $>$ 10$^{-10}$ G, as one might expect
not too far away from galaxies. In this case the 
pairs are strongly deflected and their inverse Compton
emission gives rise to a diffuse ``pair halo'' around sources,
emitting $\gamma$ rays well in excess of  10 TeV.
It is realistic to detect  ``pair halos'' with the planned
arrays of \v{C}erenkov telescopes with a wide field of view.
This would allow to determine the ``Hubble constant''
(a fundamental
cosmological parameter) in a novel way, and also to determine
the total energy output of VHE sources  in a ``calorimetric'' way,
i.e. independent of beaming effects.
\\
2. Plaga\cite{plaganat} considered the case of extremely
weak intergalactic 
magnetic fields (smaller then 
10$^{-15} G$), as they are expected far away from galaxies
as a remnant of  phase transitions 
in the very early universe\cite{siglmat}.
In this case the pairs are only slightly deflected, the ensuing
$\gamma$-rays reach the observer from the same direction as
the source, but are delayed in time due to 
the a longer travel path. These delayed events have
a distinctive distribution  in the ``delay time versus energy''
plane which allows to identify them even in the presence
of backgrounds. This 
is the only method proposed so far 
which is in principle capable to detect primordial
fields as weak as the ones expected from early-universe
phase transitions.
\\
One of the most likely extensions of the standard model
is supersymmetry\cite{deboer}. If supersymmetry
is physically realized, the lightest
supersymmetric particle (LSP) ``$\chi$'' is probably stable and 
within the mass range a 
plausible candidate for cosmological cold dark matter (CDM). 
CDM is firmly expected to gravitationally
cluster around galaxies including our
own Galaxy. The density of LSP's is largest in the
centre of our Galaxy and annihilation of LSP's
is unavoidable. In a recent paper Bergstr\"om et al.\cite{berg}
study the question of the expected intensity of
$\gamma$-ray's from annihilation products in the direction of
the Galactic centre in a general
class of SUSY models, taking into account both detailed 
calculations of the annihilation cross sections and
the expected clustering behaviour of CDM in numerical models.
The most distinctive annihilation channels are
\begin{equation}
\chi \rightarrow \gamma + \gamma 
\end{equation}
and 
\begin{equation}
\chi \rightarrow \gamma + Z_0
\end{equation}
which occur with branching ratios up to the order
of  0.1 $\%$\cite{berg}.
The resulting monochromatic $\gamma$-ray lines
at E$_{\gamma}$=M$_{\chi}$ resp. E$_{\gamma}$= 
M$_{\chi}$ (1-M$_Z^2$/4M$_{\chi}^2$)
with intrinsic widths of about 10$^{-3}$ ``have
no plausible astrophysical background whatsoever  
and would constitute a ``smoking gun'' of supersymmetric
dark matter\cite{berg}''. This is especially true of the {\it high mass}
range (m$_{\chi}>$0.6 TeV) because the astrophysical
$\gamma$-rays produce a background steeply
rising with energy (according to E$^{-2.7}$ in the canonical 
model) and the angular- and energy-resolution 
capabilities of ground based
$\gamma$ detectors improve with rising energy.
Fig.\ref{buckleyfig10} shows the comparison
of 22000 SUSY 
models considered by Bergstr\"om et al. together with
the capabilities
of existing and next generation detectors. It can be seen
that  the next generation of ground based VHE 
$\gamma$ detectors can seriously constrain the SUSY parameter
range in the high mass region and will be an important
supplement to accelerator searches for SUSY.  
\begin{figure}[tbh]
\centering
\epsfig{file=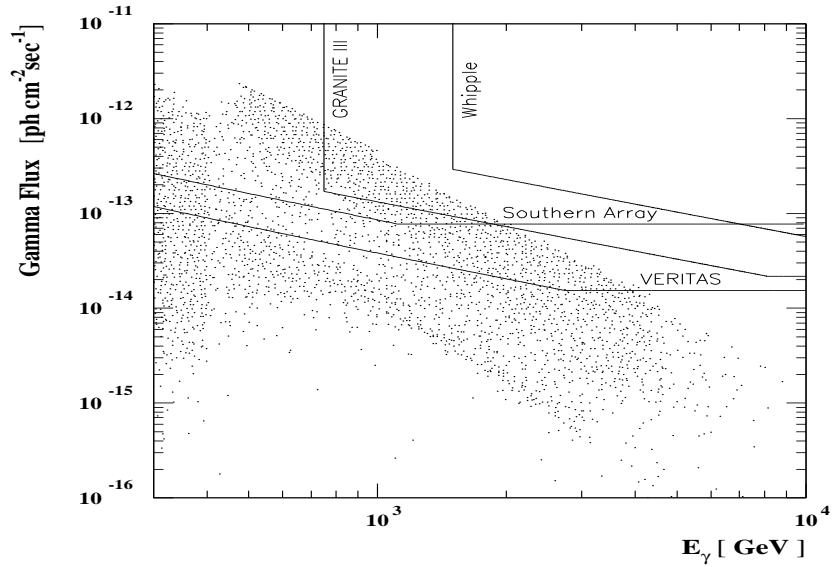,width=12cm,height=8cm,clip=,angle=0}
 \caption{\it Predicted monochromatic
$\gamma$-ray flux from a 10$^{-5}$
sr angular region centered on the Galactic center in 22000 generic
SUSY models (dots) which yield a density of LSP's in accordance 
with the required to explain CDM\cite{berg}. 
The lines are expected upper limits
from existing and planned experiments  for 10$^{6}$ seconds
observing time of the Galactic centre.
``Whipple'' is an existing single 70 m$^2$
\v{C}erenkov telescope, ``Granite III' an upgrade of this
instrument, ``VERITAS'' a planned 9 telescope array
on the northern hemisphere ``Southern Array'' stands for a similar
device (like HESS\cite{hess}), situated on the southern hemisphere of earth.  
}
\label{buckleyfig10} 
\end{figure}
Finally, extreme conditions,
that force astrophysicists to use theories
far beyond the parameter region known at the time
of their original formulation, often
prevail in the objects that can be studied with
very-high energy $\gamma$-rays. 
The history of physics makes it likely
that a more quantitative understanding of such processes, 
made possible only by using informations from {\it all} 
electromagnetic wavelength bands
and neutrinos, will 
eventually require to invoke ``new physics'' for
their description. A recent concrete proposal
in this direction is by Pen et al.\cite{pen}
In a certain double Higgs extension of the standard
model, ``$\gamma$-ray bursts''\cite{bignami} could be due a 
high-density induced baryon decay of all baryons
in a neutron star.
\section{Conclusion}
With the detection of  VHE $\gamma$-rays 
the physics of the ``nonthermal universe'' has
entered a new era. It is comparable only 
to the 1950s where the detection of nonthermal radio-emission
led to many new insights. 
\\
But will this lead to a 
resolution of some of the basic questions
in this field,
one of which has to be counted as one of the big questions 
of 20th-century physics (``Where do cosmic rays come from?'')?
Probably this will be achieved
with data from the next generation of \v{C}erenkov telescopes.  
In the inquiries about cosmic-ray origin and the mechanism of 
AGN emission the question of the relative importance of
``leptons (i.e. electrons) versus hadrons (i.e. protons and nuclei)''
remains open, however. Because
there is no ``label'' on $\gamma$-rays, indicating whether they
were produced in hadronic or leptonic reactions,
the information from  
the future detection of cosmic VHE {\it neutrinos}
in detectors like Amanda and others\cite{amanda} will
be crucial. 
\section{Acknowledgements}
I thank my colleagues form the HEGRA collaboration
for discussions and supplying material for this review.
I thank P.Biermann, M.Baring and H.V\"olk for clarifications
on theoretical issues and S.Denninghoff for a
critical reading of the manuscript.     

\end{document}